\documentclass[]{spie}  

\usepackage{listings}
\usepackage{color,fancyvrb}
\usepackage{amsmath,amsfonts,amssymb}
\usepackage{graphicx}
\usepackage[colorlinks=true, allcolors=blue]{hyperref}

\title{Automatic Spectroscopic Data Reduction using BANZAI}

\author[a,b]{Curtis McCully}
\author[a]{Matthew Daily}
\author[a,b]{G. Mirek Brandt}
\author[a,c]{Marshall C. Johnson}
\author[a]{Mark Bowman}
\author[a]{Daniel-Rolf Harbeck}
\affil[a]{Las Cumbres Observatory, 6740 Cortona Drive, Suite 102, Goleta, CA 93117-5575, USA}
\affil[b]{Department of Physics, University of California, Santa Barbara, CA 93106-9530, USA}
\affil[c]{Department of Astronomy, The Ohio State University, 4055 McPherson Laboratory, 140 West 18th Ave., Columbus, OH 43210 USA}

\begin{document} 
\maketitle

\begin{abstract}
Time domain astronomy has both increased the data volume and the urgency of data reduction in recent years. Spectra provide key insights into astrophysical phenomena but require complex reductions. Las Cumbres Observatory has six spectrographs - two low-dispersion FLOYDS instruments and four NRES high-resolution echelle spectrographs. We present an extension of the data reduction framework, BANZAI, to process spectra automatically, with no human interaction. We also present interactive tools we have developed for human vetting and improvement of the spectroscopic reduction. Tools like those presented here are essential to maximize the scientific yield from current and future time domain astronomy. 
\end{abstract}


\section{INTRODUCTION}
\label{sec:intro}  
Data reduction of spectroscopic observations has historically required significant user interaction. As our data rate increases this will not scale, either in effort or in reproducibility. Automated reduction tools for spectroscopic data are already necessary and will become even more vital as large surveys like LSST begin. To meet this growing demand, we have developed the BANZAI-NRES data pipeline to reduce data from the Las Cumbres Observatory global network of telescopes. 

\begin{figure}[!t]
\includegraphics[width=\textwidth]{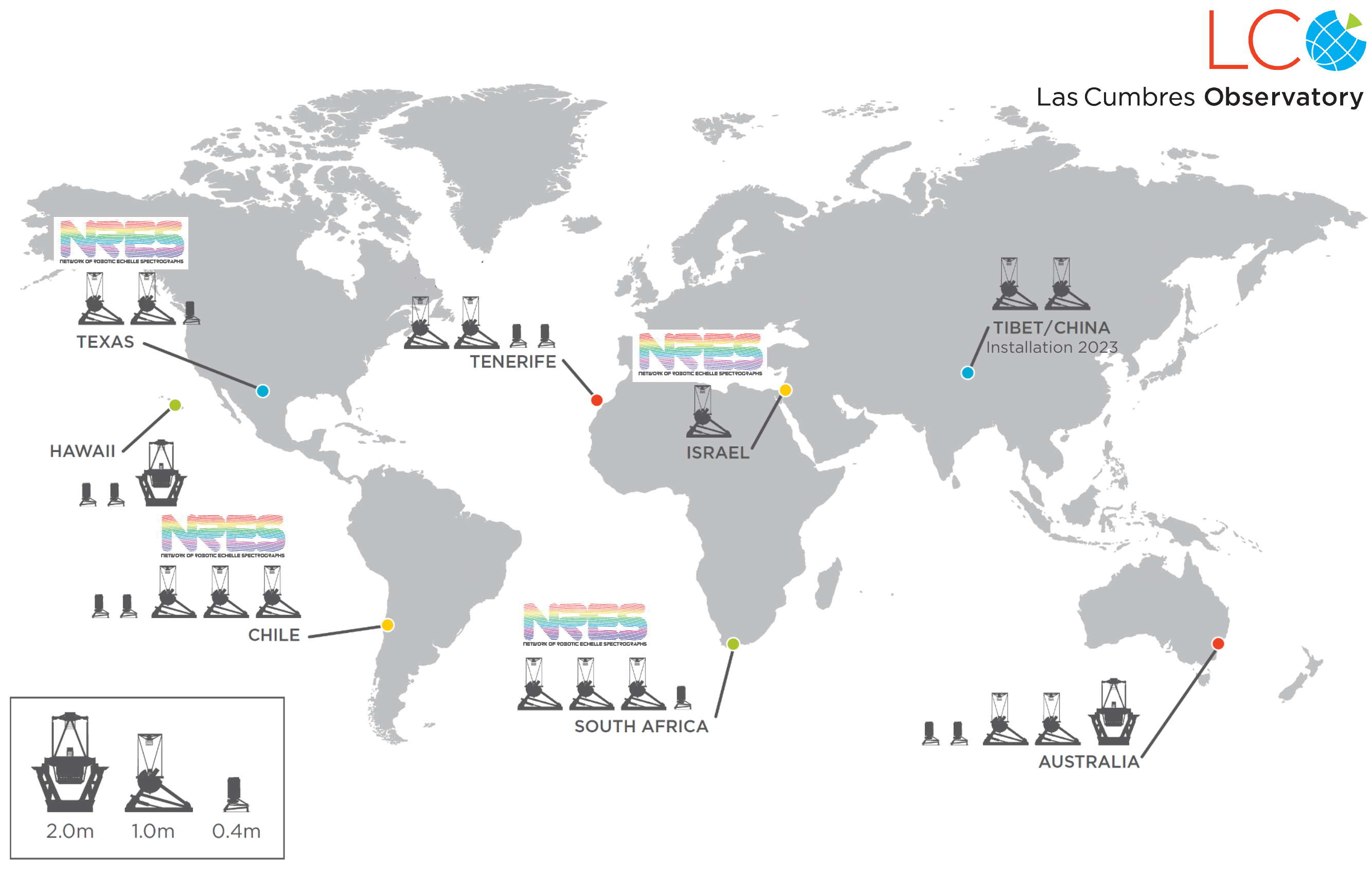}
\caption{\label{fig:lco-map}Map showing the Las Cumbres Observatory global telescope network. The sites with 1-m telescopes that are fiber-fed to NRES are marked. The FLOYDS spectrographs are at both of the 2-m telescopes.}
\end{figure}

Las Cumbres Observatory operates a network of 25 fully robotic telescopes around the world\cite{Brown2013} (see Fig. \ref{fig:lco-map}). We operate 6 spectrographs currently: the four R=50,000 units within the Network of Robotic Echelle Spectrographs (NRES)\cite{Siverd2018} and two R=500 FLOYDS instruments. Cumulatively, these instruments produce thousands of spectra every month already limiting the ability for any individual or small group to process the data manually. To meet this challenge, we developed the BANZAI pipeline that automatically detects and processes all of the data taken by Las Cumbres Observatory\cite{McCully2018}. 

BANZAI was originally developed for the Las Cumbres imaging data and is closely integrated with the observatory system: the observatory places a message on a queue (implemented via RabbitMQ: \url{https://www.rabbitmq.com/}). BANZAI listens to this queue, processing frames as they arrive. This provides both efficiency and robustness: BANZAI is not constantly checking directories and is only run on demand. The queue continues to populate even if BANZAI is down; once the pipeline is restored, it simply processes whatever is on the queue. 

BANZAI is primarily developed in Python with a small amount of C for performance. For our automated reductions, BANZAI is deployed in the cloud using Amazon Web Services (AWS). This enables rapid scalability and minimizes downtime. With the BANZAI pipeline framework in hand, we turned our attention to spectroscopic reductions beginning with the high-resolution NRES units.

Historically, the largest and best-tested suite of spectroscopic reduction tools were part of IRAF\footnote{IRAF: IRAF is distributed by the National Optical Astronomy Observatory, which is operated by the Association of Universities for Research in Astronomy (AURA) under a cooperative agreement with the National Science Foundation}. IRAF has now been deprecated and is no longer maintained (there is a community maintained distribution, but it is explicitly only for legacy projects). CERES\cite{Brahm2017} was also considered, but interoperability with BANZAI and maintaining the code base (as of this writing there have been no commits to the repo since 2019) were significant concerns. Based on this, we developed BANZAI-NRES\cite{banzai-nres} (\url{https://github.com/LCOGT/banzai-nres}, \url{https://banzai-nres.readthedocs.io/}) building on the BANZAI framework. Most RV pipelines are proprietary. BANZAI-NRES is entirely open source to meet the demand for reproducibility and collaboration. Las Cumbres is not typically a survey instrument, so BANZAI-NRES is designed to enable PI-driven science.

Our reduction methodology is presented in Section \ref{sec:methods} and performance metrics are presented in Section \ref{sec:performance}.

\section{METHODS}
\label{sec:methods}
\begin{figure}[!t]
    \centering
    \includegraphics[width=\textwidth]{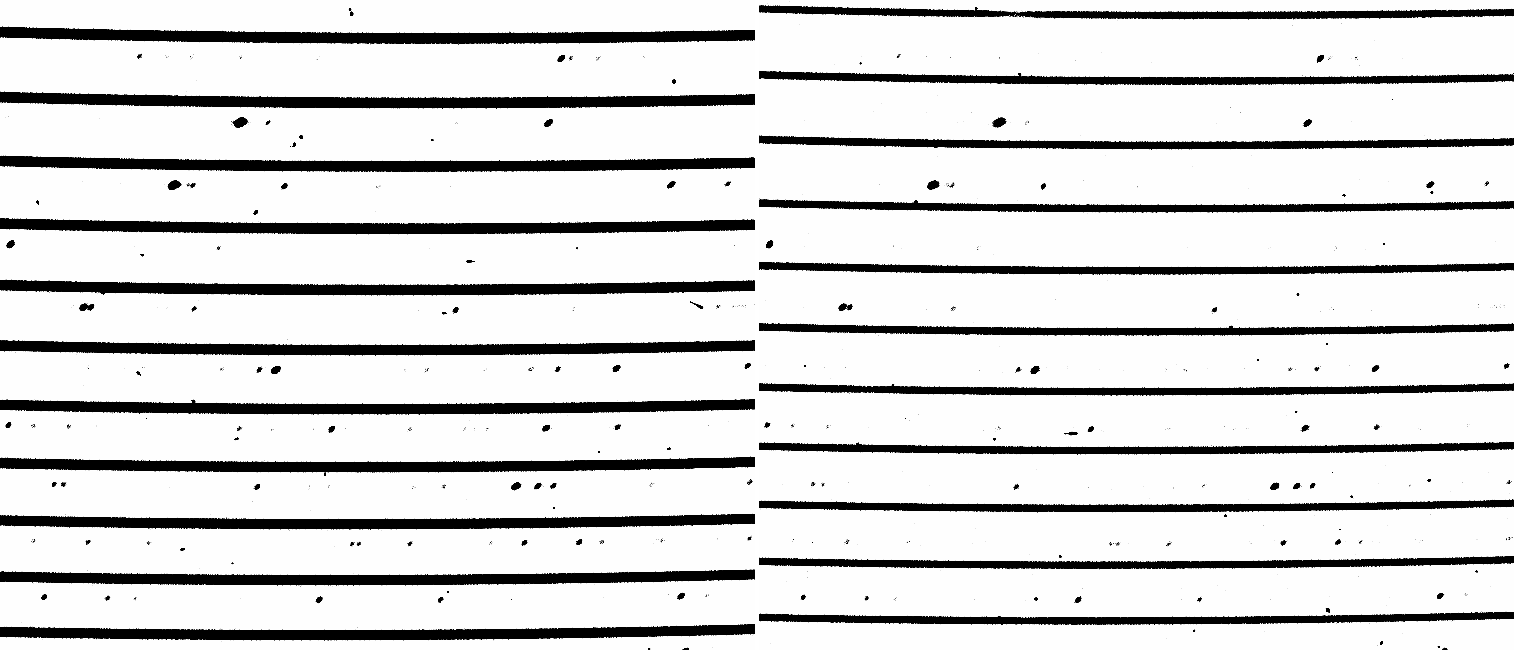}
    \caption{Science target observations with fibers lit from each telescope. The arc calibration fiber (distinct black emission dots) is lit in both left and right panels, while the projection of each science fiber (solid black traces) falls on different pixels.}
    \label{fig:nres-raw}
\end{figure}

\begin{figure}[!t]
    \centering
    \includegraphics[width=\textwidth]{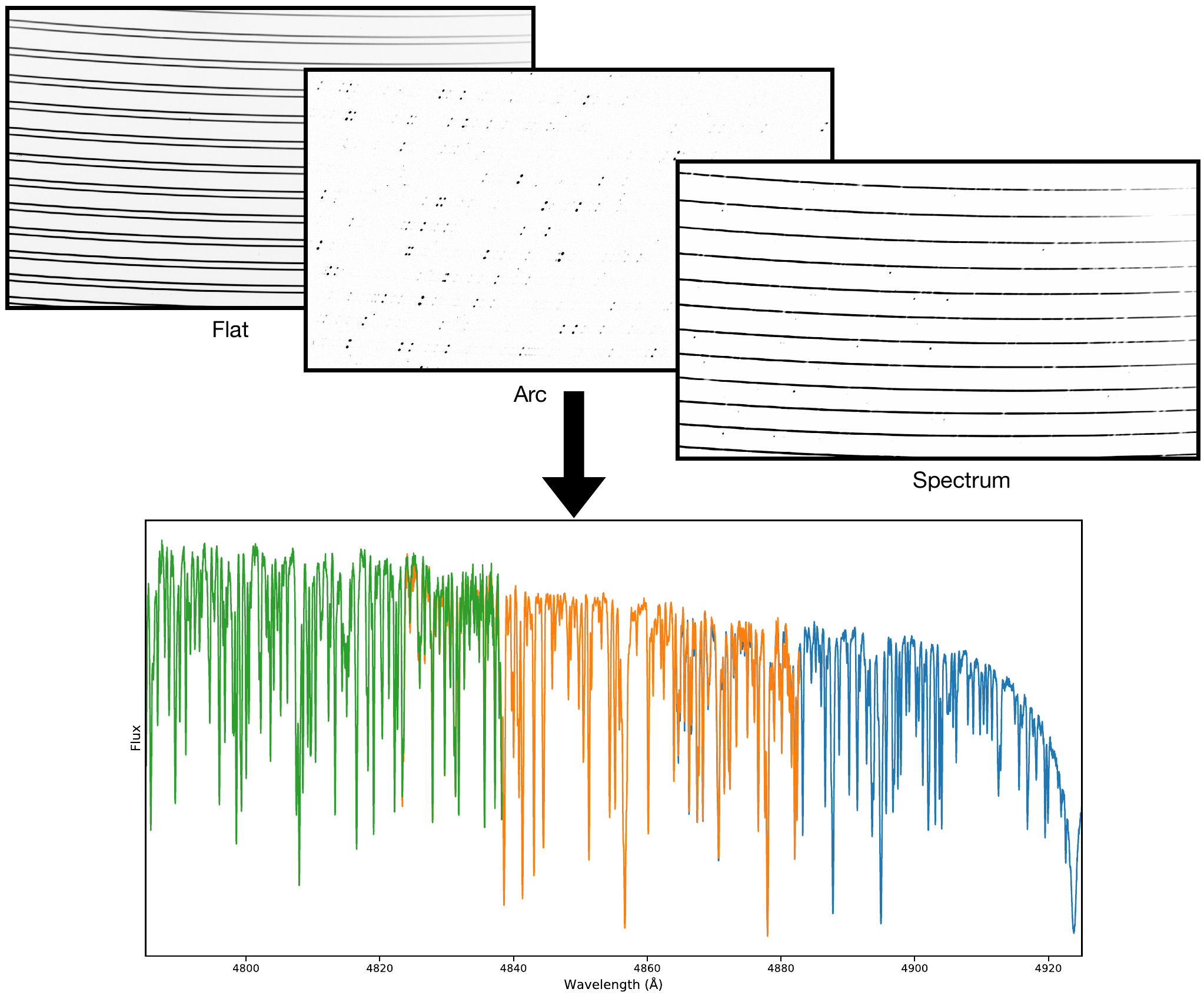}
    \caption{A schematic of the raw and processed data produced by BANZAI-NRES. This shows an example of an extraction of multiple orders. Note that no attempt at flux calibration has been made here.}
    \label{fig:extraction}
\end{figure}

\begin{figure}[!t]
    \centering
    \includegraphics[width=0.6\textwidth]{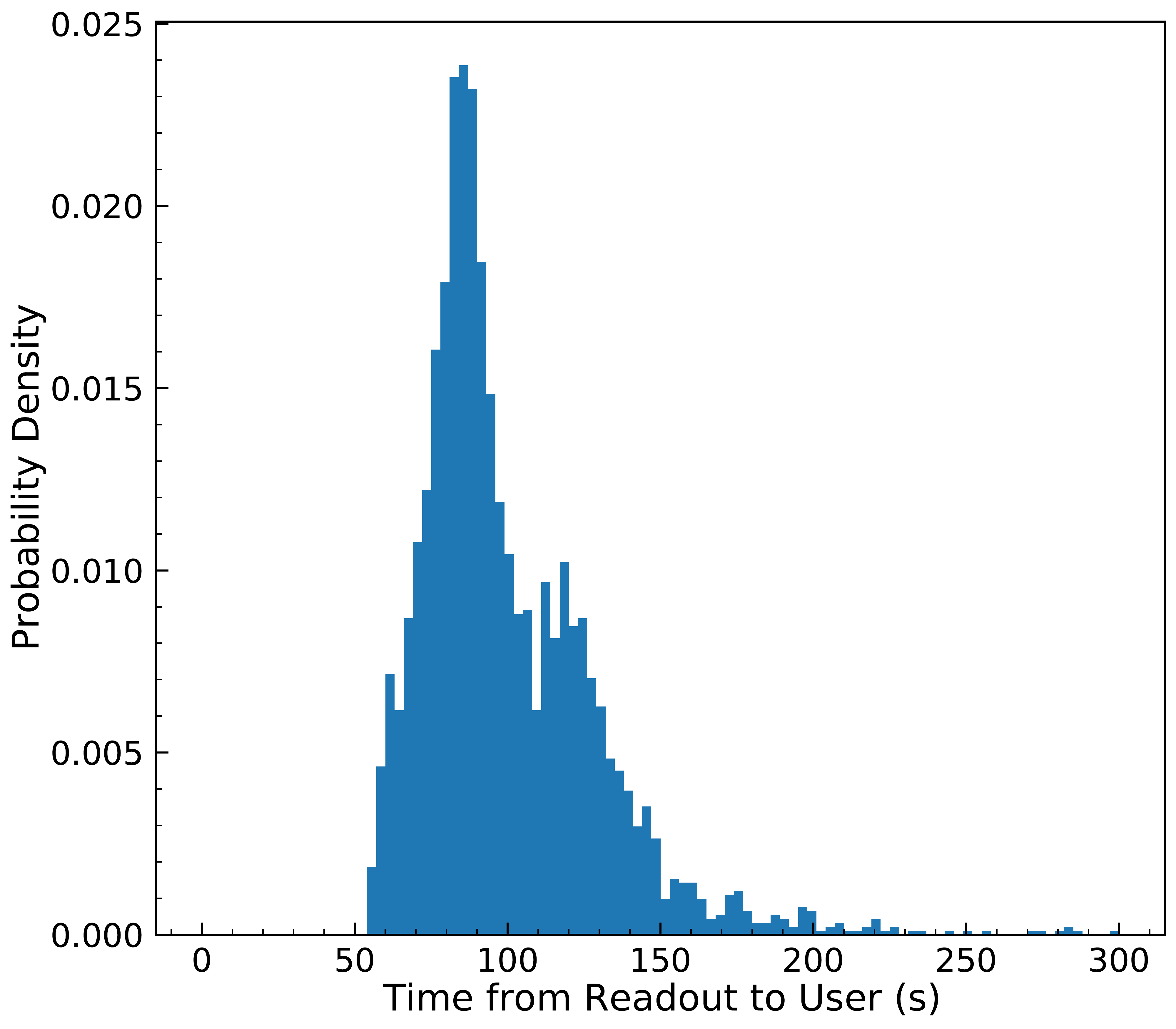}
    \caption{Distribution of time between the end of readout and reduced data available to users. The median value of this quantity is 93 seconds. This enables rapid user response which opens new opportunities as we continue in the era of time domain astronomy.}
    \label{fig:datalag}
\end{figure}

\begin{figure}[!t]
    \centering
    \includegraphics[width=\textwidth]{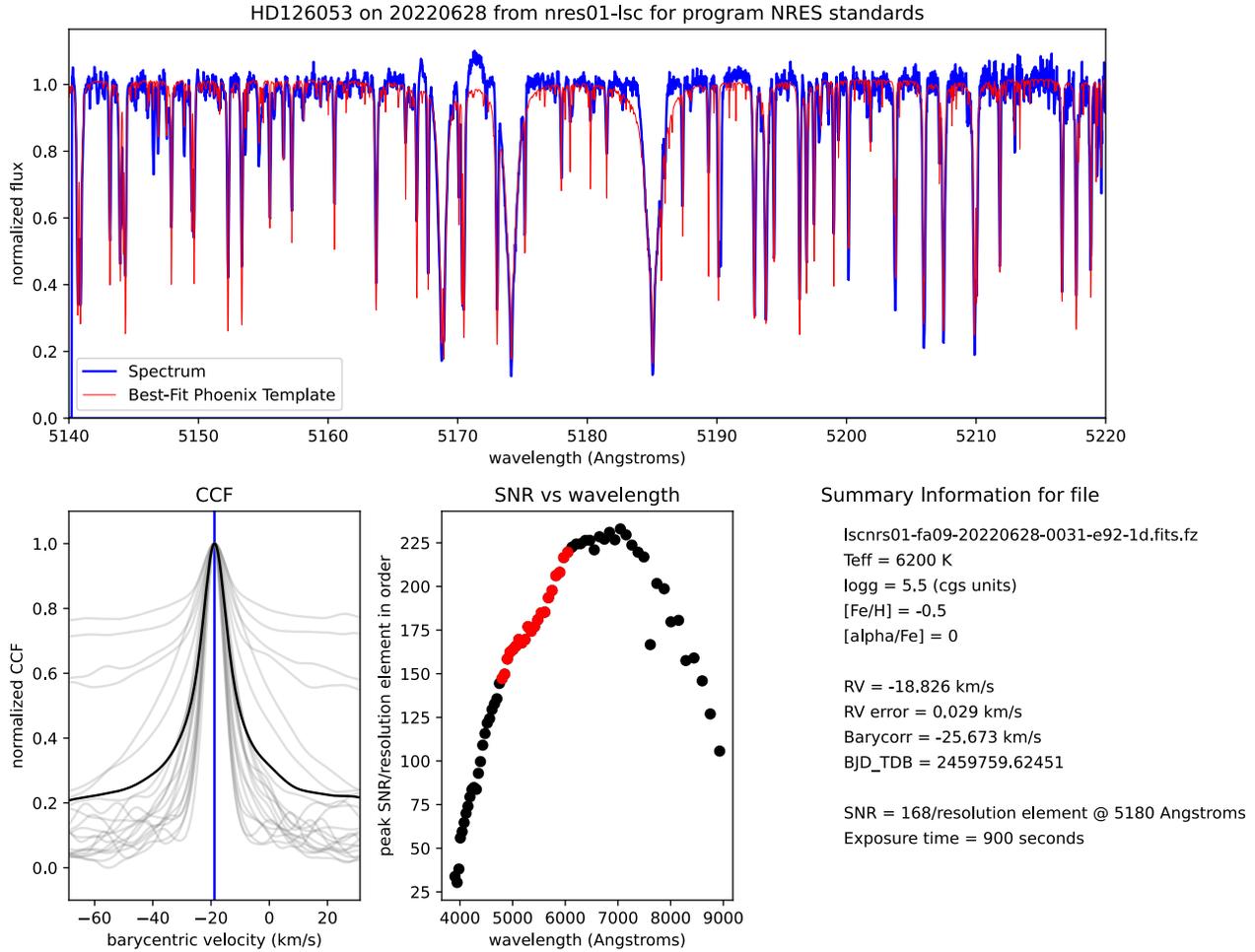}
    \caption{Summary information produced by BANZAI-NRES for user validation of their observation. The top shows the observed data in blue and the selected template in red. The template has been shifted according to the radial velocity of the target. The bottom shows the cross correlation function along with a summary of measurements on the spectrum. The information provided here allows users to quickly validate whether the signal-to-noise they expected was achieved and whether a reasonable template spectrum was chosen to infer if the radial velocity measurement is reasonable.}
    \label{fig:summary1}
\end{figure}

\begin{figure}[!t]
    \centering
    \includegraphics[width=\textwidth]{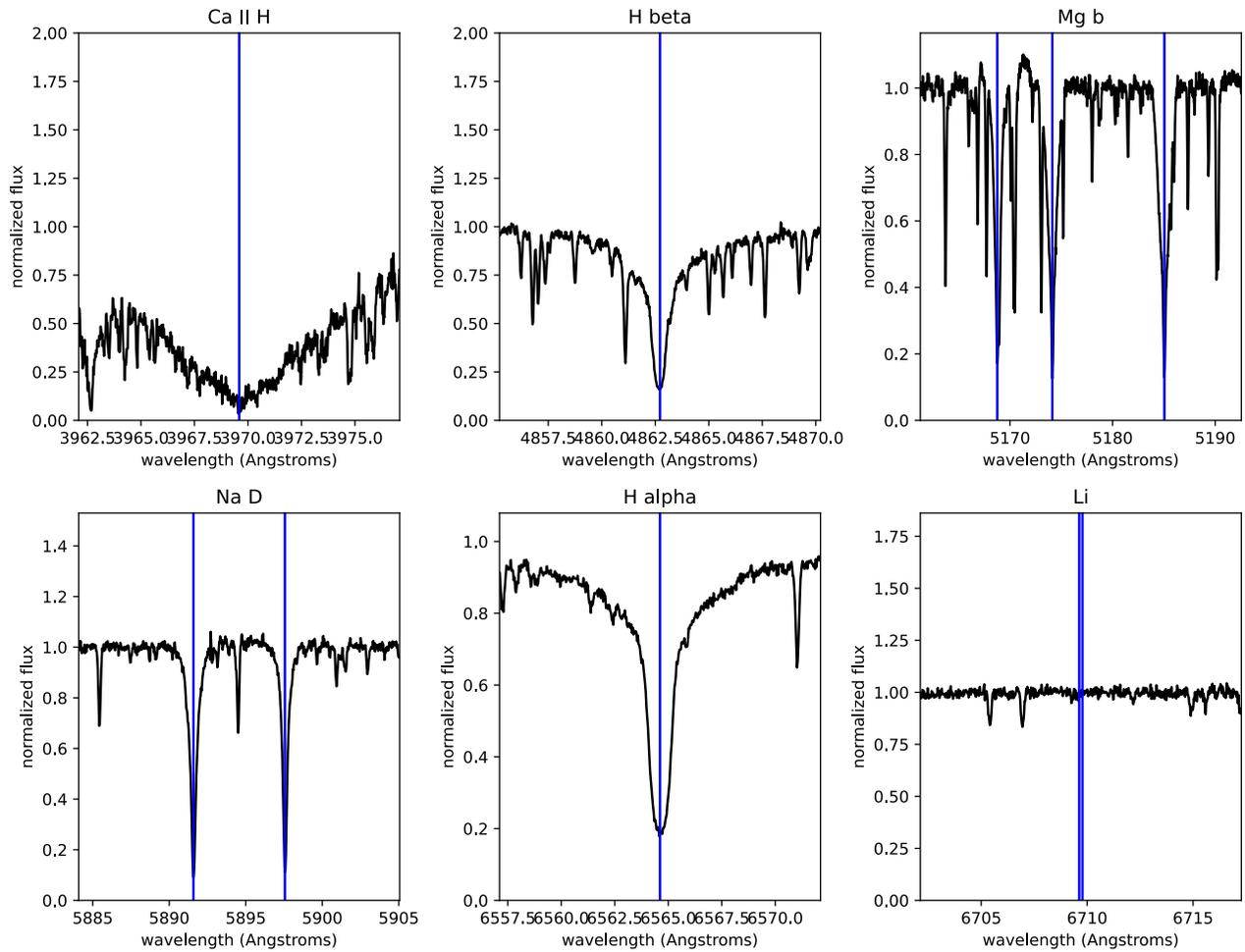}
    \caption{Section of the observed NRES spectrum near some typical strong stellar features. The extracted spectrum is shown in black while the expected central wavelength for the feature is shown with a blue line. This allows users to validate the wavelength solution and the radial velocity measurement. This summary figure along with Figure \ref{fig:summary1} and wavelengths of telluric features are provided with every processed spectrum for user vetting. This allows insight into the otherwise fully automated data reduction process.}
    \label{fig:summary2}
\end{figure}

\begin{figure}[!b]
    \centering
    \includegraphics[width=0.9\textwidth]{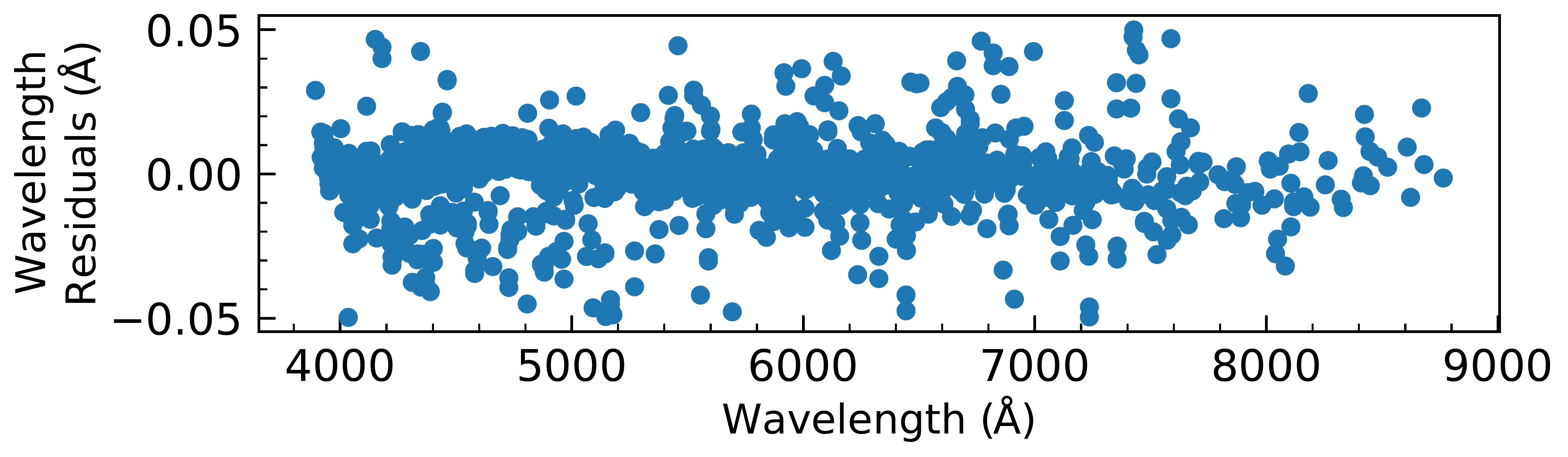}
    \caption{Wavelength residuals for a typical afternoon's set of ThAr calibration frames. The typical scatter is $<0.012$\AA. Given the $\sim1600$ lines we have on the NRES detector, this corresponds to $\sim 16.5$m/s in the radial velocity. This is roughly half of the current error bars on radial velocity measurements with NRES.}
    \label{fig:wavelength}
\end{figure}
The NRES instruments are echelle spectrographs that produce 68 dispersion orders that fall on the CCD chip. Each NRES unit is hooked to two separate telescopes (except our instrument in Israel) via an optical fiber. These two telescope fibers along with a calibration unit fiber each produce a set of orders on the chip. The calibration unit fiber is aligned such that it does not overlap the orders from either of the telescope fibers and can be used for simultaneous wavelength calibration. Figure \ref{fig:nres-raw} shows example raw 2-D frames with each pair of fibers lit. Currently, each mode is processed independently and no attempt is made to observe with both telescopes simultaneously. 

The initial processing we perform is basic CCD detector signature removal and is identical to what is used in BANZAI for imaging. We subtract a single overscan value, a full frame bias stack to remove any pattern in the readout, and a dark exposure stack to remove dark current. These stacks are produced by taking the outlier-rejected mean of each pixel. Outlier rejection is performed by comparing to the median absolute deviation (scaled to match a Gaussian of the same width for convenience). The median absolute deviation is more resilient to outliers than the standard deviation. As such, we do not need to iterate the rejection as we would if we used a traditional sigma-clipped mean algorithm. 

Throughout the data reduction, we propagate pixel-to-pixel uncertainties using analytic forms. To achieve this, we have overloaded the default arithmetic operators to include uncertainty propagation. This enables us to not include the propagation explicitly everywhere some operation is performed on the data. This maximizes code reuse, improves code readability, and minimizes the number of places bugs can be introduced. 

After removing the CCD signature, we begin the analysis specific to spectroscopy. We next detect where the orders 
land on the chip. This is a fully automated process. We begin by taking a small window of pixels near the center 
of the chip on a lamp flat field image. We then find the peaks in that window to use as the initial guess for the 
locations of each order. From those guesses, we then trace the order, stepping towards the edge of the chip. We 
encode the center of each order into a fitted Legendre polynomial. We maximize the flux in an aperture as the 
polynomial fit metric. The order locations are also stored in an integer array of the same shape as the original 
data so that users can index the data array with comparison operators 
(e.~g.~\lstinline{data['orders'] == order_number})

After operating the NRES instruments for several years, we have found that they are stable, so we only do a blind solve using the central window for major events with the instrument (e.~g.~maintenance of the camera's cryosystem). The night-to-night variations are captured by refining the polynomial fit of the order locations.

Once we have the locations of the orders on the chips, we can find the wavelength solution. We perform the wavelength solution on the 2-D frames before we extract the 1-D spectra. Our process begins with a fully automatic, blind solve for the wavelengths of each pixel. This method matches the lines in the overlapping wavelength regions between orders and is detailed in Brandt et al.~2020\cite{Brandt2020}. Likewise for retracing, we only do a full blind solve after significant instrument events. Daily ThAr arc lamp exposures are used to refine the wavelength solution; this refinement is done in the standard way by matching the nearest lines to a reference line list and fitting a Legendre polynomial. The reference line list is from taken from the UVES pipeline from ESO (\url{http://www.eso.org/sci/facilities/paranal/instruments/uves/tools/tharatlas.html}\cite{Murphy2007,deCuyper1998}). 

The orders are then extracted to produce the final 1-D spectra using the optimal extraction technique \cite{Horne1986}. Figure \ref{fig:extraction} shows an example BANZAI-NRES extraction for a few orders. The profile weights are estimated using the profile of the lamp flat field observations. An un-weighted extraction is also provided to users. These extractions are normalized to be used for radial velocities by dividing out the continuum; ``continuum'' here means a combination of the intrinsic stellar continuum, the blaze function, and the detector sensitivity function. 

If NRES has not observed this target previously, we classify the star so that we can properly choose a radial velocity template. For our automated reductions, we rely on the temperature and $
\log{g}$ estimates from Gaia\cite{Gaia2016,Gaia2018}. This catalog includes a majority of NRES targets, but is incomplete for the brightest stars. For bright stars, we query SIMBAD\cite{Simbad} for stellar parameters. Once we have an initial estimate of the temperature and $\log{g}$, we find the peak of the cross correlation function of the spectrum with the PHOENIX stellar atmosphere models\cite{Husser2013}. This classification is stored in a database and is then used for all future observations of this target so that its radial velocities are always measured with the same template. Radial velocities are measured using the same cross-correlation technique. 

All of this analysis has been fully automated and requires no human intervention. The algorithms were chosen to ensure they are robust against a variety of failure modes. Thanks to their robustness, the algorithms we have chosen have a few initial parameters that need to be set carefully making much of BANZAI-NRES broadly applicable to a range of echelle spectrographs. For instance, the wavelength calibration routine has been shown to work on some HARPS spectra\cite{Brandt2020}.

Robotically scheduling telescopes is more efficient than giving out classical nights to individual PIs, but the data volume is such that no individual person could visually inspect all the data. This means that we must detect issues programatically and it is even more important than normal to produce high quality, bug-free code. To this goal, we have adopted a testing pattern of defining unit tests for each stage of the pipeline on synthetic data. As we know the ground-truth from the input data, we can characterize each stage individually. We also include end-to-end testing of BANZAI-NRES on real data to ensure that there are no issues plumbing the stages together, the stages indeed work on real data from the telescopes, and the quality of the reduction does not get worse than previous versions. Performance metrics for BANZAI-NRES are presented below in Section \ref{sec:performance}.

\section{PERFORMANCE}
\label{sec:performance}

\subsection{Data}
BANZAI-NRES was deployed to production in May 2021 and has reprocessed to the beginning of the 2021A semester. So far, BANZAI-NRES has processed more than 8500 science target spectra. Modern time domain astronomy requires a rapid turnaround taking the data from the telescope, processing it, and serving it to users. Figure \ref{fig:datalag} shows the distribution of the time between the end of readout on the telescope and availability to users. The typical frame, including processed and wavelength-calibrated science-ready spectra, is served to its requester $93$ seconds from readout.

As every spectrum is processed without user intervention, it is necessary to provide users with visualizations for quality control. Figures \ref{fig:summary1} and \ref{fig:summary2} show example summary plots for a given observation. Figure \ref{fig:summary1} shows the template matched to the observed spectrum validating the choice of radial velocity template. Several other summary measurements are provided here as well. Figure \ref{fig:summary2} shows some of the typically strongest features in the spectrum to validate that the wavelength solution is correct and that the radial velocity measurement is reasonable. These summary figures are provided to the users alongside the reduced data to enable rapid validation of the processing.

\subsection{Wavelength Calibration}

\begin{figure}[!t]
    \centering
    \includegraphics[width=0.8\textwidth]{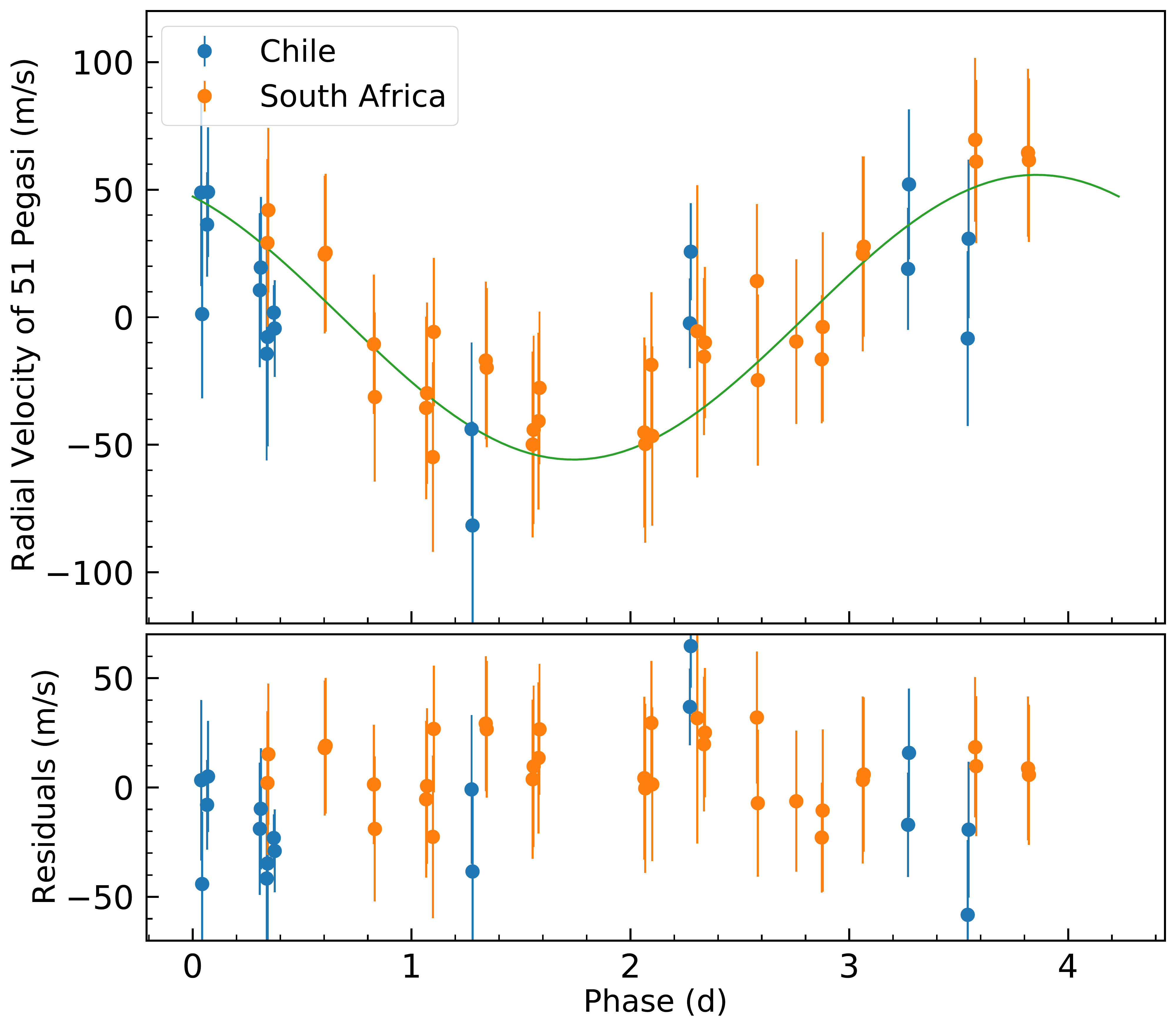}
    \caption{Radial velocity curve of 51 Pegasi taken with NRES instruments at two different sites. The data is phase folded to the orbital period. We have overplotted a simple zero-eccentricity model using the previously measured period\cite{Mayor1995}. Our measurements using the automated BANZAI-NRES pipeline are in agreement with theirs. Our median error bar that we calculate by analytic uncertainty propagation for this data set is 32 m/s. This is a consistent representation of the data: the RMSE we estimate from the known radial velocity curve of 23 m/s. This factor of 1.4 overestimate of the uncertainties is likely due to covariance that is not included in our analytic models. }
    \label{fig:rv}
\end{figure}

The fundamental metric for a radial velocity instrument with the resolution of NRES that sets the floor on the radial velocity precision for solar type stars is the quality of the wavelength solution. The NRES hardware is stable enough that we can use the wavelength solution based on ThAr arc lamp exposures taken the previous afternoon. Figure \ref{fig:wavelength} shows the typical quality of a wavelength solution from BANZAI-NRES. The points correspond to individual lines measured in the extracted spectra that are in the reference calibration line list. The residuals have an RMS of $\sim 0.01$\AA. This corresponds to a velocity precision of $\sim 16.5$~m/s for the $\sim 1600$ lines in the ThAr spectrum. 

There are a few planned, upcoming improvements to the wavelength solution that will improve this performance. The first issue we have identified is that there is some astigmatism away from the center of the chip. We can model this by doing a full 2-D wavelength solution and then including that in 2-D information in the extraction as opposed to extracting straight along columns. A second improvement we have planned is to use the simultaneous calibration fiber arc lamps to track the variation of the fiber positions over the night. This is not currently a limiting factor in the overall wavelength solution quality but as we implement the 2-D wavelength solution we expect that it will be. These improvements will bring us closer to our original goal of 3 m/s radial velocity precision.

\subsection{Radial Velocities}

The original science case for building the NRES instruments was to efficiently measure radial velocities for bright stars ($V \gtrsim 10$). While NRES are valuable as general use echelle spectrographs, radial velocity measurements are typically the most demanding on performance and so are a good metric on the quality of our data processing. Figure \ref{fig:rv} shows the NRES-measured radial velocity curve of 51 Pegasi, the host of a hot Jupiter\cite{Mayor1995}. The line shows a simple zero-eccentricity model with the previously measured period and RV semi-amplitude. The NRES data are in good agreement with previous measurements from other instruments. The residuals have an RMS of $23.4$ m/s which we use as an estimate of the radial velocity precision. Comparing to our analytic estimations of the uncertainties, we find that BANZAI-NRES overestimates the radial velocity uncertainty (or conversely underestimates the actual precision) by a factor of $1.4$. This is likely due to covariance that is not included in our uncertainty models.

Beyond the improvements to the wavelength solution discussed above, there are improvements that can be made that will increase the radial velocity precision of the automatic BANZAI-NRES reductions. These include improved background subtraction to remove scattered light, telluric correction, and corrections for fringing. We expect each of these to contribute small improvements to our overall radial velocity precision.

\section{CONCLUSIONS}

BANZAI-NRES is a completely automatic data reduction pipeline to analyze high-resolution echelle spectra and to produce radial velocity measurements of bright stars. This pipeline has already processed more than 8500 target spectra and will continue to provide high quality reductions to Las Cumbres Observatory users. As these tools must be robust to enable complete automation, they are broadly applicable to a wide range of echelle spectrographs. We are currently adapting these to work with our FLOYDS spectrographs at Las Cumbres Observatory. Some modifications are necessary because FLOYDS has a slit while NRES is a fiber. Working with FLOYDS data also requires us to disentangle multiple objects (e.~g.~a galaxy host and a supernova). Once we complete this, BANZAI will include a set of spectroscopic processing tools that are applicable to a wide range of spectrographs. We aim to repackage these tools into a set that is easy to apply to new instruments so that they can be used by the wider community. 

\acknowledgments   
 
We thank Tim Brown for the initial NRES concept, implementing the commissioning data pipeline, and many discussions about processing NRES data. We thank Markus Rabus for enlightening discussions about NRES and data processing. We thank Lisa Storrie-Lombardi for her support and management of the project.

\bibliography{report} 
\bibliographystyle{spiebib} 

\end{document}